\documentstyle[11pt,newpasp,twoside]{article}
\reversemarginpar
\begin{document}
\title{Magnetic Fields of Neutron Stars: an Overview}
 \author{Andreas Reisenegger}
\affil{Departamento de Astronom\'\i a y Astrof\'\i sica, Pontificia 
Universidad Cat\'olica de Chile, Av. Vicu\~na Mackenna 4860, Santiago, Chile}

\begin{abstract}
Observations indicate that magnetic fields on neutron stars span 
at least the range $10^{8-15}$ G, corresponding to a range of magnetic
fluxes similar to that found in white dwarfs and main sequence stars.
The observational evidence is discussed, as well as the possible
origin of the field, and the associated phenomenology
(``classical'', millisecond, and binary pulsars, ``magnetars'', etc.). 
Particular attention is given to physical processes potentially 
leading to magnetic field evolution.
\end{abstract}

\section{Introduction}

Neutron stars are dense, compact remnants of evolved stars, with
degenerate fermions compressed by strong gravitational fields (e.g., 
Shapiro \& Teukolsky 1983). Their structure
is set almost entirely by one parameter, their mass. The latter can be measured
with some accuracy only in binary systems containing a pulsar, where
its distribution is found to be consistent with a narrow Gaussian,
centered at $1.35 M_\odot$ and with width $0.04 M_\odot$ (Thorsett \& 
Chakrabarty 1999), so one might presume that neutron stars are essentially 
all identical. This is far from true. A wide range of rotation rates and 
magnetic fields, 
together with the presence or absence of mass-transfering binary companions, 
allow for a rich phenomenology. Among single (or non-accreting binary) 
neutron stars, we distinguish ``classical'' pulsars, millisecond pulsars,
soft gamma-ray repeaters, anomalous x-ray pulsars, and inactive, thermal 
x-ray emitters. Binary systems with mass transfer onto a neutron star can 
be divided into high-mass and low-mass x-ray binaries (according to the 
companion mass), with substantially different properties. 
Magnetic fields play an essential role by accelerating particles,
by channeling these particles or accretion flows, by producing synchrotron 
emission or resonant cyclotron scattering, and by providing the main
mechanism for angular momentum loss from non-accreting stars. It is even
speculated that in some objects the magnetic field may be the main energy 
source for the observed radiation.

I begin by reviewing several observational ``classes'' of neutron
stars mentioned above, with a special eye on the evidence for the 
presence and strength of the magnetic fields in each class (\S 2). In \S 3,
the resulting magnetic fluxes are compared to those of other kinds of
stars, and possible connections are discussed. Section 4 surveys the
evidence for and against magnetic field evolution, and discusses
physical processes which may lead to such evolution. General conclusions
are presented in \S 5.

\section{Classes of neutron stars and evidence for magnetic fields}

\subsection{Radio pulsars}

Radio pulsars are regularly pulsating sources of radio waves, 
interpreted as magnetized, rotating neutron stars (Pacini 1967; Gold 1968). 
Beams of radiation emerging from 
the poles of a roughly dipolar magnetic field misaligned with respect to the
rotation axis appear as pulses every time they sweep the location of the 
Earth. These pulses reveal rotation periods ($P$) from 1.55 milliseconds (ms) 
to several seconds, which lengthen with time ($\dot P>0$). 

The simplest model for the spin-down process is to consider the neutron star
as a magnetized body of moment of inertia $I$, rotating in vacuum with angular 
velocity $\vec\Omega$ (Ostriker \& Gunn 1969). It 
loses rotational energy due to the time-variation of
its magnetic dipole vector $\vec\mu$, which is rotating at a fixed
inclination $\alpha$ with respect to to the rotation axis,

\begin{equation}
-{d\over dt}\left({1\over 2}I\Omega^2\right)={2\over 3 c^3}|\ddot{\vec\mu}|^2
={1\over 6c^3}B^2R^6\Omega^4\sin^2\alpha,
\end{equation}

\noindent which allows us to infer the dipole magnetic field strength,
$B[{\rm G}]\approx 3.2\times 10^{19}\sqrt{P[{\rm s}]\dot P}$ (for radius
$R=10$ km, $I=10^{45}{\rm g\,cm}^2$, and $\alpha=90^\circ$).

This ``dipole
in vacuum'' model is unlikely to be very accurate, as real pulsars are
surrounded by a magnetosphere, and by an interstellar medium whose plasma
frequency is much higher than the expected radiation, which therefore
can't propagate. Somewhat more realistic models (e.g., Gold\-reich
\& Julian 1969) tend to roughly confirm the estimate of $B$ 
(being less sensitive to $\alpha$), so this estimate is generally used.
Assuming a constant field strength and moment
of inertia, eq. (1) can be integrated backwards in time to give
a divergent rotation rate at a time $\tau=\Omega/(-2\dot\Omega)=P/(2\dot P)$
before the present (at which the spin parameters are to be evaluated), 
defining a characteristic ``spin-down age'' for the pulsar.

In terms of these parameters, radio pulsars fall into two fairly disjoint
groups (e.g., Phinney \& Kulkarni 1994):

\begin{itemize}
\item young ($\tau\sim 10^{3-7}$ yr), relatively slow ($P\sim 16$ ms 
to several seconds), and strongly magnetized 
($B\sim 10^{11-13}$ G) ``classical'' pulsars, and

\item old ($10^{8-10}$ yr), fast (1.55 to several ms), and weakly
magnetized ($10^{8-9}$ G) ``millisecond'' pulsars.
\end{itemize}

Confirming that $\tau$ is related to true age, many of the ``youngest'' 
classical pulsars are found to be associated with supernova remnants
(which disperse after $\sim 10^5$ yr), and many millisecond pulsars
(but no classical pulsars) are found in globular clusters. An additional
difference between the two classes is that most millisecond pulsars
are found in binary systems (in most cases with old white dwarf companions), 
whereas the vast majority of classical pulsars are single.

One problem with this general picture is that,
in the few cases where it is possible to measure $\ddot\Omega$
(all of which are young pulsars), the so-called braking index 
$n\equiv\Omega\ddot\Omega/\dot\Omega^2$ does {\it not} agree with
the canonical $n=3$ predicted in the dipole spin-down model, but has
significantly {\it smaller} values, which differ from one pulsar to another
(e.g., Kaspi et al. 1994). 
The inclusion of higher multipoles (e.g., quadrupole electromagnetic or
gravitational radiation) worsens the problem. 
This means that the {\it inferred} magnetic dipole moment in young pulsars
{\it increases} with time. Whether this corresponds to a true increase
of the star's dipole moment has not been settled (Blandford 1994).

\subsection{Magnetars}

Two puzzling kinds of astronomical objects have in recent years found a
likely interpretation as very highly magnetized neutron stars 
(see Thompson 2000 for a review):

\begin{itemize}
\item Soft gamma-ray repeaters (SGRs)
are a class of (so far) 4 objects which repeatedly emit bursts of gamma-rays, 
in addition to persistent x-rays. For two of these sources, regular
pulses have been observed in the persistent x-ray emission, allowing the 
measurement of a rotation period and period derivative (Kouveliotou et al. 1998;
Hurley et al. 1999). 

\item Anomalous x-ray pulsars (AXPs) show persistent x-ray
emission, modulated at a stable, slowly lengthening period.
Contrary to the standard, {\it binary} x-ray pulsars (\S 2.3), 
they show no evidence 
for a companion star (see Mereghetti \& Stella 1995; 
van Paradijs, Taam, \& van den Heuvel 1995; Mereghetti 2000).  
\end{itemize}

Aside from the presence of bursts in the SGRs, these two classes of objects
appear to be very similar. All measured periods lie in the narrow range
$5-12$ s, and objects in both classes are associated with supernova remnants
(e.g., Kaspi 2000),
arguing for an interpretation as young neutron stars, in rough agreement 
with characteristic ages inferred from spin-down. 
In both classes of objects, the persistent x-ray luminosity is much larger
than the inferred spin-down power. Therefore, unlike the case of radio pulsars,
rotation can {\it not} be a significant energy source. The dipole fields inferred
from the spin-down rate are $10^{14-15}$ G, much larger than in previously
known classical radio pulsars, though pulsars with similar inferred dipole
fields have recently been found (Camilo et al. 2000). It has long been suggested
that magnetic energy may be the ultimate source of both the bursts and
the persistent radiation (Duncan \& Thompson 1992; Paczynski 1992; 
Thompson \& Duncan 1995, 1996), 
but this would still require a total magnetic energy
significantly larger than inferred from the dipole field, i.e. a buried 
and/or disordered magnetic flux. In any case, the strong magnetic field may 
modify the radiation transport in the surface layers, so that these objects
radiate a much larger fraction of their fossil heat in x-rays (as opposed to
neutrinos) than less magnetic neutron stars (Heyl \& Hernquist 1997a, b).

\subsection{X-ray binaries}

Neutron stars accreting from binary companions are a vast field of research
on their own, which I cannot possibly cover here. I only
point out that neutron stars with high-mass companions tend to appear as
x-ray pulsars, in which the accreted material is channeled by the magnetic
field onto the polar caps, whereas low-mass companions tend to live with
non-pulsating neutron stars, in which the field is presumably not strong
enough to channel the accretion flow. In some members of the first class,
cyclotron features have been found in the x-ray spectrum, corresponding
to $B\sim (1-4)\times 10^{12}$ G (e.g., Makishima et al. 1999). 
Note that these are the only direct 
measurements of neutron star magnetic fields, akin to the many measurements
of magnetic fields on white dwarfs and other stars. Assuming that these
objects differ from (similarly young) classical radio pulsars only by
the presence of the nearby companion, this would give evidence that the
field of neutron stars is organized on a relatively large scale, so the 
surface field and the dipole field are of comparable magnitude.

\section{Kinship}

Most, if not all, neutron stars descend from main sequence stars with
masses $M_{MS}\ga 8 M_\odot$, i.e., O and early B stars, while lower mass
main sequence stars give rise to white dwarfs.
We have learned that a fraction of early-type stars (Ap/Bp stars) have strong, 
highly organized magnetic fields (see the presentations of G. Mathys, 
J.D. Landstreet, S. Bagnulo, and N. Piskunov in this volume). The same is
true for a fraction of the white dwarfs, which tend to be more massive
than their non-magnetic counterparts (G. Schmidt, this volume), and 
therefore plausibly more closely related to neutron stars.

It has long been known that the magnetic fluxes of magnetic white dwarfs
and neutron stars are similar (e.g., Ruderman 1972), suggesting a common 
origin, possibly through flux conservation during the evolution from 
some progenitor phase. Although much more strongly magnetic
objects have been discovered in recent years, the most strongly magnetic
main sequence stars (Ap/Bp stars with $R\sim$ few $R_\odot$ and 
$B\sim 3\times 10^4$ G; e.g. Landstreet 1992), white dwarfs 
($R\sim 10^{-2}R_\odot$, $B\sim 10^9$ G; e.g., Wickramasinghe \& Ferrario 
2000), and neutron stars (magnetars with $R\sim 10^{-5}R_\odot$ and
$B\sim 10^{15}$ G; see \S 2.2) still turn out to have remarkably similar
magnetic fluxes, $\Phi=\pi R^2B\sim 10^{5.5}R_\odot^2$ G, despite vast 
differences in size, density, and magnetic field strength. Lower limits
on magnetic fluxes can unfortunately not be compared, as the magnetic fields
of most non-degenerate stars and white dwarfs are too weak to be detected.
Of course, we may also not yet know the most magnetic stars, if they are
scarce or manifest themselves phenomenologically in a way we have not yet
identified.

Another interesting point is as follows. Early-type stars have convective
cores and radiative envelopes. The mass of the convective core, $M_{conv}$, 
is a strongly increasing function of the total mass of the star, $M_{MS}$
(e.g., Kippenhahn \& Weigert 1994). The mass of the eventual
compact remnant, $M_{rem}$, is a much more weakly increasing function of 
$M_{MS}$ (e.g., Weidemann 1987). The two curves cross at
$M_{MS}\approx 3-4 M_\odot$ ($\sim$ A0 stars), where
$M_{conv}=M_{rem}\approx 0.7 M_\odot$, a plausible dividing line between
magnetic (massive) and non-magnetic (low-mass) white dwarfs (cf. G. Schmidt,
this volume). A possible interpretation is that during the main 
sequence phase the field exists only in the convective core of most 
early-type stars. (A coherent, equipartition-strength field filling the
convective core of a $4M_\odot$ main sequence star produces approximately 
the maximum flux estimated above.) If it remains confined to the same region 
during the later stages of evolution, then low-mass white dwarfs have 
a magnetized region buried in their interior and covered by an unmagnetized
envelope (formerly part of the radiative envelope on the main sequence),
whereas massive white dwarfs and neutron stars form exclusively from 
magnetized material, and therefore have a strong surface field.

\section{Magnetic field evolution}

\subsection{Observational evidence}

Several arguments point toward the possibility of an evolving magnetic field
in neutron stars:

1) Generally speaking, young neutron stars appear to have strong magnetic fields
$\sim 10^{11-15}$ G (``classical'' radio pulsars, ``magnetars'', x-ray pulsars),
whereas old neutron stars have weak fields $\la 10^9$ G (ms pulsars,
low-mass x-ray binaries). If these two groups have an evolutionary connection,
their dipole moment must decay. Millisecond pulsars are believed to
have been spun up to their fast rotation by accretion from a binary
companion, a remnant of which is in most cases still present (e.g., Phinney 
\& Kulkarni 1994). Accretion may be the direct or indirect 
cause of the reduction in the magnetic dipole moment, or it may just be
an effect of age.

2) Studies of the pulsar distribution on the $P-\dot P$ diagram (analogous
to ``normal'' stellar population synthesis studies on the HR diagram)
have led to the claim that the magnetic torque decays on a time scale
comparable to the life span of ``classical'' pulsars (Gunn \& Ostriker 1970).
The case for this claim was strengthened by the simultaneous consideration
of pulsar space velocities and their spatial distribution perpendicular to
the plane of the Galaxy (e.g., Narayan \& Ostriker 1990), but was later
put in doubt by other authors (e.g., Bhattacharya et al. 1992),
whose more careful analysis leads to opposite results. 

3) If magnetar emission is powered by magnetic energy (Thompson \& Duncan 
1996), then the rms magnetic field $\langle \vec B^2\rangle^{1/2}$ must decay.  

4) A possible explanation for the ``anomalous'' braking indices $n<3$ in
young neutron stars is that their magnetic dipole moment increases with
time.

In the remainder of this section, I discuss the physical mechanisms that
may lead to such an evolution of the magnetic field.  

\subsection{Physics of spontaneous field evolution}

The composition of neutron star matter is still
highly uncertain (e.g., Lattimer \& Prakash 2000), but it seems almost
inevitable that it will contain both neutral particles (plausibly neutrons)
and charged particles (protons, electrons, and possibly others). All 
particles are highly degenerate. The relativistic energies of the electrons 
reduce their cross-section for colliding against protons, and most of
the phase space for final states is blocked by the Pauli principle, leading 
to a high conductivity and 
consequently to an Ohmic decay time longer than the age of the Universe
(Baym, Pethick, \& Pines 1969b). Therefore, little diffusion of the
magnetic field can occur.

Can the magnetic field move {\it with} the fluid matter inside the neutron
star, driven by magnetic stresses or buoyancy forces? Not in an obvious
way. The matter in an equilibrium neutron star is fully catalyzed,
i.e., weak interactions have had time to bring each fluid element into
chemical equilibrium, minimizing its free energy by distributing baryon
number optimally among different ``flavors'' of particles. This optimal
distribution is density-dependent, giving rise to a {\it mechanically
stable composition gradient} (Pethick 1992; Reisenegger \& Gold\-reich 1992), 
regardless of the uncertainties in the
composition (Reisenegger 2001). Even in the simplest and most favorable
scenario, in which the matter is mostly neutrons, with a small ($\sim 1\%$) 
``impurity'' of protons and electrons, magnetic stresses of order the
``impurity'' contribution to the fluid stresses are required to 
overcome the stabilizing forces, demanding a magnetic field $\ga 10^{17}$ G.
At lower field strengths, the magnetic stresses can only build up a small
chemical imbalance, and evolve on a timescale determined
by the weak interactions which erase this imbalance.

This leads to ask whether the magnetic field could move only with
the charged particles, leaving the neutral particles behind. This question
was addressed in a simple model (Gold\-reich \& Reisenegger 1992) in which 
protons and electrons move under the effect of electromagnetic forces through 
a static and uniform neutral background, scattering against each other and 
against this background. It leads to the following evolution law for the 
magnetic field,

\begin{equation}
{\partial\vec B\over\partial t}=\nabla\times\left(\vec v\times\vec B\right)
+\gamma\nabla\times\left(-{\vec j\over n_e e}\times\vec B\right)
-\nabla\times\left({c\over\sigma}\vec j\right),
\end{equation}

\noindent where $\vec v$ is a weighted average velocity of all charged
particles, $\vec j=c\nabla\times\vec B/(4\pi)$
is the electric current (due to {\it relative} motions of the charged 
particles), $n_e$ is the density of protons and electrons, $e$ is the proton 
charge, $c$ is the speed of light, $\sigma$ is an {\it isotropic
conductivity}, limited by inter-particle collisions, and $\gamma$ is a
dimensionless factor ($|\gamma|<1$) whose magnitude and sign depends on
the relative coupling of protons and electrons to the neutral
background. Each term on the right-hand side has a familiar (astro-)physical 
interpretation, in turn:

1) Advection of the magnetic flux by a {\it flow of charged particles,}
i.e., {\it ambipolar diffusion,} familiar from star formation (e.g.,
T. Mouschovias, this volume): The bulk flow arises from magnetic stresses
or buoyancy forces acting on the charged particles, and is impeded by 
inter-particle collisions. It can be decomposed into two modes, one
curl-free and one divergence-free, the first of which will be choked by
the chemical potential gradients it builds up in the charged particles, and 
can only be effective if weak interactions can reduce these gradients.
Since the driving forces are $\propto B^2,$ this term is $\propto B^3,$
becoming much more effective at high field strengths.

2) Advection of the magnetic flux by the electric current, or {\it Hall
effect:} 
This is a ``passive'' or ``kinematic'' effect, not ``driven'' by any forces and
which by itself does not change the magnetic energy. However, it is 
nonlinear ($\propto B^2$) and could possibly lead to small-scale structures
in the magnetic field, particularly in the solid crust, where ambipolar 
diffusion cannot occur.

3) The familiar {\it resistive} or {\it Ohmic diffusion}: Linear in $B$;
it is quite ineffective for a large-scale field, but may play a role in 
dissipating small-scale structures created by the other (nonlinear)
processes.

For the flows disturbing chemical equilibrium (bulk flow
and curl-free ambipolar diffusion), the timescale is set by weak interactions,
which also produce the early cooling of neutron stars (through neutrino 
emission), and which are strongly temperature dependent. Therefore, if
$B\la 10^{17}$ G, the only way in which these processes can be effective 
before the star cools down is to keep it hot by some other mechanism,
such as dissipation of magnetic energy (e.g., Thompson \& Duncan 1996).
However, even this is not guaranteed to work, since most of the dissipated
energy will be emitted in the form of neutrinos. If the
field is strong enough to create a substantial chemical imbalance 
($B\ga 10^{16}$ G), the enhanced neutrino emission may even lead to
faster cooling.

None of these mechanisms appear to be interesting at field
strengths and time scales relevant to classical or millisecond radio
pulsars, unless the magnetic field is confined to a thin layer 
in the outer crust of the star, where the conductivity 
is reduced and a combination of Hall drift and Ohmic dissipation may 
become effective. In magnetars, the high field strength makes
both the Hall drift and the ambipolar diffusion quite fast, and their
interaction may lead to interesting dynamics, particularly if the rms 
interior field is somewhat higher than the inferred dipole field, as
required from energetic arguments.

This discussion did not consider the formation of superfluid and superconducting states, which probably
occurs early in the evolution of a neutron star (Baym, Pethick, \& Pines 
1969a), concentrating vorticity and magnetic flux into quantized ropes. 
At moderate to low temperatures, the neutron star fluid will be much more 
complicated than in the description given above (Mendell 1998). 
The effect of these complications on magnetic field evolution
are not yet well-understood, though much has been speculated. I will refrain
from further discussion of these issues. 

\subsection{Induced field evolution}

Aside from the mechanisms for spontaneous evolution, external agents may
also change the magnetic field of a neutron star:

1) The strong thermal gradient in a cooling protoneutron star is able
to overcome the stratifying effect of the chemical gradient, leading to
convection. At the same time, the star has not had time to transport
angular momentum and will be differentially rotating. As discussed several 
times in this conference, this combination
naturally acts as a dynamo, which is an alternative to the ``fossil flux'' 
idea to give rise to the magnetic field in neutron stars (Thompson \& 
Duncan 1993).

2) The thermal gradient persists for a much longer time in the outer crust
of the star, where it may act as a battery, again giving rise to a substantial
field (Urpin \& Yakovlev 1980; Blandford, Applegate, \& Hernquist 1983; Wiebicke 
\& Geppert 1996 and references therein). This may in
principle explain an increasing field in a young neutron star, as suggested
by the braking index measurements.

3) Accretion from a binary companion is a natural (and popular) way
of decreasing the magnetic dipole moment, although there is no agreement
on the exact physics involved. Perhaps the most interesting candidate process
is the burial of the magnetic flux by the accreted, highly
conducting plasma (Bisnovatyi-Kogan \& Komberg 1975; Romani 1993). 
No full models of this process have been produced so far, and 
three-dimensional simulations will eventually be needed to make sure all
possible instabilities have been taken into account.
If effective, this process
still begs the question of why after its completion a minute, but fairly 
constant fraction of the initial dipole moment is left or regenerated
to be detectable in ms pulsars. (Note that ms pulsars do {\it not} appear to
be the
``tip of the iceberg'' of a distribution extending down to much lower fields,
since the death rate of their progenitor low-mass x-ray binaries already can 
barely account for the detectable ms pulsars; e.g., Phinney \& Kulkarni 
1994; White \& Ghosh 1998.)

\section{Conclusions}

The subject of magnetic fields in neutron stars
has still much to offer and to demand from us. Little is known about the
structure of the field, but nevertheless there is a rich phenomenology asking 
to be interpreted. The origin and evolution of the field is as uncertain as
in all other kinds of stars. Interesting physics is at play, and there may be 
connections between the magnetic fields of neutron stars, their white dwarf 
cousins, and the main-sequence progenitors of both. It is likely that advances
in the understanding of the respective magnetic fields will support each other 
if sufficient communication is maintained among the communities of experts,
as occurred so well in the present meeting.

\acknowledgments

The author is grateful to Peter Gold\-reich for extensive conversations
and many insights which have benefited the present review. He also thanks 
Lars Bildsten for related discussions and for his hospitality at the
Institute for Theoretical Physics in Santa Barbara, where a previous 
version of this presentation was prepared. The writing of this article was
supported by FONDECYT (Chile) under grant no. 8970009 ({\it L\'\i neas 
Complementarias}).

\newpage

\section*{Discussion}

\noindent VAN BALLEGOOIJEN: What is the frequency of the radiation responsible
for the energy loss from pulsars? What are the physical processes involved?

\noindent REISENEGGER: Taking the magnetic dipole spin-down model seriously,
we expect the radiation to be nearly monochromatic at twice the
rotation frequency of the star. This is far below the
plasma frequency of the interstellar medium, so this radiation
cannot propagate. In practice, the spin-down appears to occur through
more complicated processes which are not really understood, but almost
certainly involve the acceleration of charged particles and the resulting
synchrotron and ``curvature'' radiation (the latter being due to relativistic
particles moving along curved magnetic field lines). I don't think there is
a theoretical prediction for the resulting spectrum. Observationally, the 
spectral energy distribution ($\nu F_\nu$) from pulsars 
has a wide peak at high energies (GeV gamma-rays to hard x-rays), 
where most of the energy is emitted. It falls towards lower 
energies, from the ultraviolet to the radio.
The bolometric radiation output, although uncertain due to incomplete 
frequency coverage and beaming effects, appears to vary from a small fraction
of the pulsar spin-down power in the most energetic pulsars to a substantial 
fraction in less energetic objects. The remainder of the energy is carried
away by a wind and eventually deposited in the surrounding ``plerion'' nebula. 

\noindent STRASSMEIER: Does the polarisation of the radio signal (or any 
other radiation that you may detect) tell something about the magnetic field?

\noindent STAIRS: The polarization can tell us about the geometry of the
magnetic field in the emitting region, but not about its magnitude. 
The latter is obtained exclusively from timing of the pulsar spin-down. 

\noindent HENRICHS: Could you comment on alignment between the magnetic
and rotation axis as a possible mechanism to cause the apparent decay of the
field?

\noindent REISENEGGER: In the standard dipole spin-down model, this is
clearly a possibility, as we are only sensitive to the dipole component 
perpendicular to the rotation axis. This has been explored several times, 
most recently by Tauris and Manchester. 
Theoretically, it doesn't appear much easier to
align the dipole than to make it disappear, though people are clever and
do come up with ideas. In addition, slightly more
sophisticated models for the spin-down process, such as the early one
by Gold\-reich and Julian, predict spin-down even when the dipole is
aligned with the rotation axis (the only case considered in that particular
work), which seems to imply that the spin-down process is not very sensitive
to the orientation of the magnetic dipole.

\noindent SCHMIDT: What fraction of neutron
stars is magnetic? In other words, how does the pulsar birth rate compare
to the core-collapse supernova rate?

\noindent REISENEGGER: That ratio is hard to nail down, as both the numerator
and the denominator have large error bars which are difficult to quantify.
From statistics alone, we cannot safely discard, on the one hand, that all
supernovae produce pulsars or, on the other, that only a modest fraction do.
However, it is interesting to note that many young supernova remnants do not
contain detectable radio pulsars but rather x-ray sources (AXPs, SGRs, 
``quiescent'' neutron stars) or no compact source whatsoever. 

\end{document}